\documentclass[%
 reprint,
 amsmath,amssymb,
 aps,
floatfix,
]{revtex4-2}

\usepackage{graphicx}
\usepackage{dcolumn}
\usepackage{bm}
\usepackage{subcaption}
\usepackage{float}
\usepackage[capitalise]{cleveref}
\usepackage{chemformula}

\begin{document}

\preprint{APS/123-QED}

\title{The impact of steric repulsion on the capacitance and total free energy of electrochemical systems}

\author{Dagmawi Berhanu Tadesse}
\email{dagmawi.tadesse@murdoch.edu.au}
\affiliation{Discipline of Physics,  Chemistry and Mathematics, C'SHEE, Murdoch University, 90 South St, Murdoch, 6150 WA, Australia}
\author{Drew Francis Parsons}%
 \email{drew.parsons@unica.it}
\affiliation{Department of Chemical and Geological Sciences,
University of Cagliari, Cittadella Universitaria, 09042 Monserrato, CA}

\date{\today}

\begin{abstract}
We present a modified Poisson-Boltzmann model that adds steric repulsion of ions, providing an approximation to ion-ion correlations. Controlled by specific ion sizes, this steric contribution provides a concentration cap.  Importantly, the steric contribution to the total free energy is comparable in size to the electrostatic contribution and exceeds the classical energy $\frac{1}{2}CV^2$. Capacitance varies nonmonotonically with voltage, with a maximum near 0.2--0.3V depending on the ion size. An analytic expression for the total free energy is derived in terms of the size of the counterion, surface potential, and charge density at each electrode.
\end{abstract}
\maketitle

The Poisson-Boltzmann (PB) model is a  mean field theory which treats ions in solution as point-like particles. The solvent is treated as a continuous medium represented by a dielectric constant. With ion size unaccounted for, this model grossly overpredicts the concentration of counterions near the surface of charged particles for concentrations as low as 20 mM and electric potentials as weak as few tens of mVs \cite{Bai2007a,Stellwagen2011a}. As a result PB theory predicts an unphysically unbounded capacitance of the electric double layer. The model also neglects short range ion-ion interactions which are vital in determining the concentration profiles and the total free energy of the electric double layer.

Stern was the first to recognize the limitation of point-like ions in 1924 \cite{Stern}. Since then there have been numerous efforts made to account for ion size, both within the frame work of mean field theory and in the statistical mechanics of particles on a lattice \cite{Bikerman1942,Eigen1954,Carnahan1969a,Boublik1970,Mansoori1971a,Kralj-Iglic1994,Iglic1996,Bohinc2002,Borukhov1997a,Borukhov2000a,Borukhov2004a}. More recently Borukhov \cite{Borukhov1997a} derived expressions of the concentration profile of finite sized ions for both symmetric and asymmetric valences. Kilic \cite{PhysRevE.75.021502} and Kornyshev \cite{Kornyshev2007a} came up with an expression for charge density and capacitance of symmetric valence electrolytes and equal size of anion and cation. While Bazant, Storey, and Kornyshev \cite{PhysRevLett.106.046102,Goodwin2017a} introduced short range ion-ion interactions for equal sized anions and cations, Han \cite{Han2014a} extended on \cite{PhysRevE.75.021502,Kornyshev2007a} to account for electrolytes with asymmetric valence electrolytes. Gupta and Stone \cite{Gupta2018a} studied the effect of valence asymmetry on  dielectric value of the solvent, and
ion-ion correlations. More recently, Pedro de Souza \cite{PhysRevLett.125.116001} proposed a free-energy functional to describe the coupling between the long range electrostatic forces and the short range ion-ion interactions. 

In this letter,  we present the impact on total free energy from steric interactions. We add a  steric interaction term to the excess chemical potential of ions which is activated when a certain energy threshold is surpassed. This added term prevents overcrowding of ions, similar to what hard-sphere ion-ion correlations would achieve, and can therefore  be taken as an approximate representation of short range ion-ion correlations. Importantly, our approach invokes individual ion sizes, providing a framework for understand Hofmeister (specific ion) effects in electrochemical systems \cite{Wojciechowski2011,KasuyaSogawaMasudaKamijoUosakiKurihara2016}.  

Consider two parallel flat-surfaced electrodes held at a potential difference $V$ immersed in electrolyte solution. The chemical potential of the $i^{\text{th}}$ ion at any position in the solution with bulk concentration $c_{i\infty}$ is given by

\begin{equation}\label{chemPot}
  \mu_i(x) = \mu^{\text{en}}_i(x) + \mu^{\text{ex}}_i(x)
\end{equation}

\noindent where $\mu^{\text{en}}_i = k_BT\ln(\chi_i)$ is the entropic chemical potential,  $k_B$ is Boltzmann constant, $T$ is temperature and $\chi_i$ is mole fraction. $\mu^{\text{ex}}_i = \mu^{\text{el}}_i + \mu^{\text{st}}_i$ is the excess chemical potential, in which we include  the electrostatic component  $\mu^{\text{el}}_i$ and, importantly, a steric interaction $\mu^{\text{st}}_i$. $\mu^{\text{el}}_i=z_i\mathrm{e}\Phi$, where  $z_i$ is the valence number of the $i^{\text{th}}$ ion, $\mathrm{e}$ is the elementary charge, $\Phi$ is the electrostatic potential. The steric interaction $\mu^{\text{st}}_i$ is a short range interaction which is activated when an ion is in close proximity of another ion preventing ion overlap. The expression for the steric interaction can be derived by setting the respective long range electrostatic force, $F^{\mathrm{el}}=- \nabla\mu^{\mathrm{el}}_i$, and the local steric force, $F^{\mathrm{st}}=-\nabla\mu^{\mathrm{st}}_i$, be equal and opposite, $F^{\mathrm{st}}(\Phi) = -F^{\mathrm{el}}(\Phi)$. Integrating this expression to find an expression for the steric chemical potential as

\begin{eqnarray}
   \mu^{\mathrm{st}}_i &=&  \mu^0_i - \mu^{\mathrm{el}}_i 
\end{eqnarray}
where $\mu^0_i$ is an integration constant defined to be $\mu^0_i=- k_BT\ln\left(c^{\mathrm{cap}}_i/c_{i\infty}\right)\equiv\mu^{\mathrm{cap}}_i$, which is the steric energy threshold that corresponds to a concentration cap controlled by how far the ions can get in proximity to one another. In this paper, for simplicity we used hard-shphere of the ions as the distance of closest approach. Hence $c^{\mathrm{cap}}_i \sim 1/V_i$. $c_{i\infty}$ is the bulk concentration. Thus the steric term is turned on when the threshold, $\mu^{\mathrm{cap}}_i$ is surpassed and it is given as \cite{Borah2011a,Parsons2016a} 
\begin{equation}
    \mu^{\mathrm{st}}_i= 
\begin{cases}
    \mu^{\mathrm{cap}}_i -\mu^{\mathrm{el}}_i ,& \mathrm{if }~ \mu^{\mathrm{el}}_i < \mu^{\mathrm{cap}}_i\\
    0,              & \mathrm{otherwise}
\end{cases}
\end{equation}
In thermal equilibrium, for a chemical potential given by Eq.\eqref{chemPot}, the Poisson-Boltzmann equation will have the form
\begin{equation}\label{MPB}  
    -\nabla^2 \Phi(x)
		=\sum_{i}^{N}\frac{N_{\mathrm{A}}\mathrm{e}z_ic_{i\infty}}{\varepsilon_0\varepsilon_s}\exp\left(-\frac{\Delta\mu^{\mathrm{ex}}_i}{k_BT}\right)
\end{equation}
\noindent where $N_{\mathrm{A}}$ is Avogadro's number, $\varepsilon_0$ is vacuum permitivity, $\varepsilon_s$ is the dielectric function of the solvent medium, $k_B$ is Boltzmann constant. The total free energy of the system accounts for the contributions of all components of  the  chemical  potential. From Eq.\eqref{chemPot}, the total free energy will follow as \cite{Parsons2012a,Gray2018a}
\begin{eqnarray}
  \nonumber  F &=& \sum_i\int dx~ \mu_i(x)N_{\mathrm{A}}c_i(x)  \\
    F &=& F_{\mathrm{en}} + F_{\mathrm{el}} +F_{\mathrm{st}} 
\end{eqnarray}
\noindent where $F_{\mathrm{en}}$ is the entropic contribution given by \cite{Theodoor1990}
\begin{equation}\label{eq:Fen}
    F_{\mathrm{en}} = k_BTN_{\mathrm{A}}\sum_{i}\int dx\left[c_i(x)\ln \frac{c_i(x)}{c_{i\infty}}-c_i(x) + c_{i\infty}\right]
\end{equation}
\noindent $F_{\mathrm{el}}$ is the electrostatic contribution given by 
\begin{equation}\label{eq:Fel}
    F_{\mathrm{el}} = \frac{\varepsilon_0\varepsilon_s}{2}\int dx \left|\nabla\Phi(x)\right|^2
\end{equation}
\noindent and the steric contribution, $F_{\mathrm{st}}$, given by
\begin{equation}\label{eq:Fst}
    F_{\mathrm{st}} = \sum_{i}\int dx ~\mu_i^{\mathrm{st}}(x)N_{\mathrm{A}}c_i(x)
\end{equation}
Our model treats the steric interaction as a direct ion interaction that reacts in opposition to other ion interaction potentials, and represents repulsive overlap of ionic electron clouds. This stands in contrast to an excluded-volume approach \cite{Bikerman1942,Paunov1996a,Borukhov1997a,Sugioka2016,PhysRevE.75.021503,PhysRevLett.125.188004}, which achieved capped ion concentrations indirectly through the addition of solvent entropy to $F_{\mathrm{en}}$, in place of the direct steric term $F_{\mathrm{st}}$.

We apply the model to five different salt solutions, \ch{LiCl}, \ch{LiPF6}, \ch{LiC2F6NO4S2} (or simply \ch{LiTFSI}), 1-Ethyl-3-methylimidazolium bis(trifluoromethylsulfonyl)imide (or simply \ch{EMTFSI}) and \ch{LiX} where \ch{X-} is an idealized anion with a maximal ion size chosen to be large enough such that its concentration is capped at its bulk concentration.   Motivated by energy storage applications, the solvent is taken as propylene carbonate solvent, of dielectric constant $\varepsilon_s = 66.14$ and  bulk ion  concentrations taken as $c_{\infty}=1.2\mathrm{M}$.

Eq.(\ref{MPB}) is solved numerically by Finite Element methods using FEniCS \cite{AlnaesBlechta2015a}.  The separation between the electrodes is $20\lambda_D$,  chosen to be large enough that the double layer interaction between the electrodes is negligible.   $\lambda_D=(\varepsilon_0\varepsilon_s k_BT/\sum_i z_i^2\mathrm{e}^2c_{i\infty})^{1/2}$ is the Debye-length. 

\begin{table}[h]
\begin{ruledtabular}
\begin{tabular}{lccr}
\textrm{Ions}&
\textrm{$R_i (\textup{\AA})$}&
\textrm{$c^{\mathrm{cap}}_i$ (M)}&
\textrm{$\Phi_{ci}$ (mV)}\\
\colrule
\footnote{Hard-sphere radii evaluated using radii of Gaussian spatial distributions of ions \cite{Parsons2009a,Parsons2014c}.}\ch{Li+} & 0.42 & 5350 & -216  \\
\footnote{Sizes estimated from ion geometries using Avogadro\cite{Hanwell2012}.}\ch{EM+} & 3.82 & 7.1 & -46 \\
$^{\text{a}}$\ch{Cl-} & 2.05 & 46 & 94 \\
$^{\text{a}}$\ch{PF6-} & 2.54 & 24 & 77 \\
$^{\text{b}}$\ch{TFSI-} & 3.76 & 7.5 & 47\\
\ch{X-} & 6.91 & 1.2 & 0 \\
\end{tabular}
\end{ruledtabular}
\caption{Ion properties: hard-sphere radius $R_i$, concentration cap $c^{\mathrm{cap}}_i$ and steric threshold potential $\Phi_{ci}$ at 298K. \ch{X-} is an idealized anion with size chosen such that its concentration cap equals  bulk concentration 1.2M. }
\label{table}
\end{table}

\begin{figure}[ht]
\centering
    \includegraphics[width=0.5\linewidth]{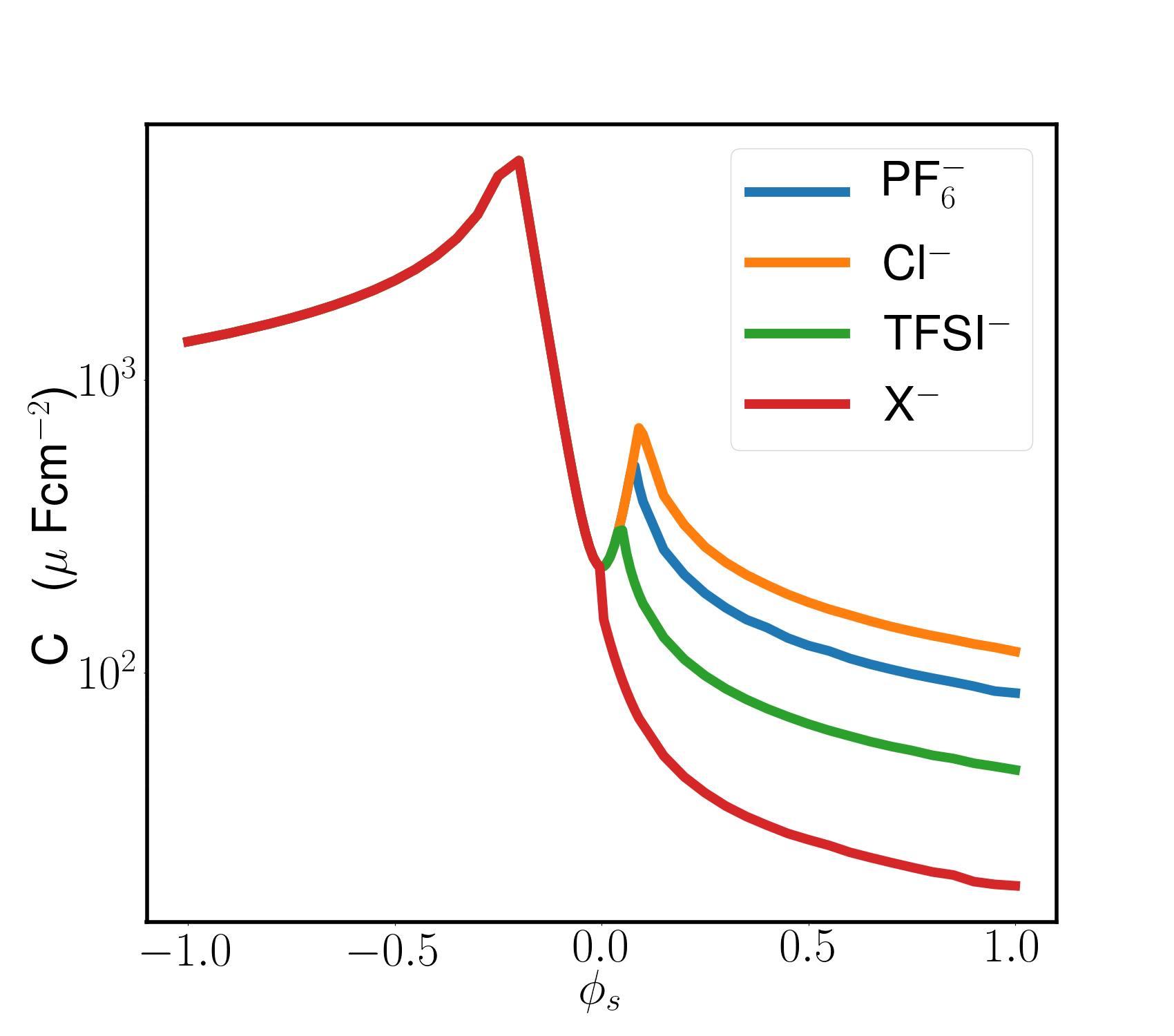}\hfil
    \includegraphics[width=0.5\linewidth]{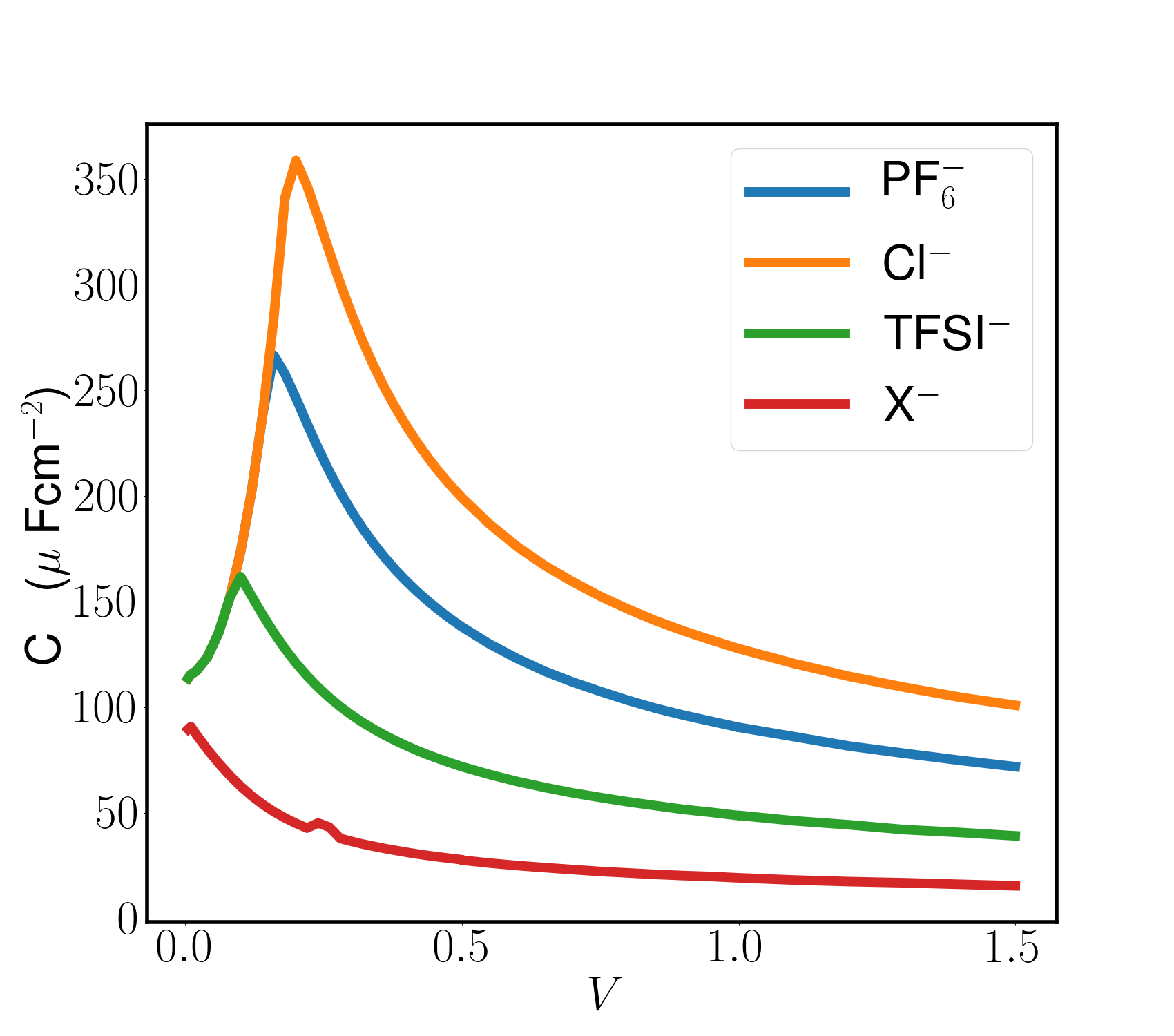}
\caption{(Left) Differential capacitance of a single electrode versus surface potential (Right) Total differential capacitance of two electrodes versus potential difference between the two electrode with \ch{Li+} as a cation and \ch{PF6-}, \ch{Cl-}, \ch{TFSI-} and \ch{X-} as anions in both cases.}\label{Fig:Caps}
\end{figure}

The differential capacitance at the electrode is computed from the surface charge density, $\sigma= -\varepsilon_0\varepsilon_s \nabla\Phi|_{x=0}$, as $C_\alpha=d\sigma_\alpha/d\Phi_{s\alpha}$. $C_{\alpha}$ is evaluated numerically using $d\Phi_{s\alpha} = 0.1$ mV, where $\alpha = +,-$. The total differential capacitance is calculated as ${1}/{C} = \sum_\alpha {1}/{C_\alpha} $. For a single electrode, as can be seen in Fig.\ref{Fig:Caps} (Left), the capacitance is independent of the co-ion size, which is indicated in the figure having identical values for negative potential. For a two electrode system, the size of the co-ions have an effect on the capacitance of each electrodes owing to the conservation of electrode charges. Charge conservation across the electrodes must apply, with the total sum of the charge densities over all electrodes being zero. Due to the size difference between anion and cation, this charge neutrality condition is achieved through an asymmetric split of the potential difference between the two electrodes \cite{Parsons2014b}. $(\Phi_{s\alpha} \ne V/2)$, where $\Phi_{s\alpha}$ is the electrode potential. Thus the total differential capacitance controlled by the bigger ions. The differential capacitance, shown in Fig.\ref{Fig:Caps}, matches that of the composite diffuse-layer capacitance in Ref\cite{PhysRevE.75.021502}. Beyond the steric threshold, triggered by potential differences around 100--400 mV, the capacitance is ion specific (controlled by ion size) and  decays as a power function of the potential, $C\propto \Phi_{s\alpha}^{-0.5}$ \cite{Bazant2009a}. 

From \crefrange{eq:Fen}{eq:Fst}, the total free energy is a function of the applied potential, size of the ions and bulk concentration. We investigated the potential and size dependence of the total free energy and compared it against $E=\frac{1}{2}CV^2$, which is conventional calculation of the energy of such system, as shown in Fig. \ref{Fig:energyComp} and \ref{Fig:EneVsPD} below.

For potentials lower than the steric  threshold potential  $\Phi_{ci} = \mu^{\mathrm{cap}}_i/z_i\mathrm{e}$ (typically 50--200 mV, see Table~\ref{table}), only the entropic and electrostatic components contribute to the total free energy. As shown in Fig.\ref{Fig:energyComp}, in this low potential regime, irrespective of the bulk concentration, the total free energy is approximately equal to $\frac{1}{2}CV^2$, where $C$ and $V$ are the total capacitance and potential difference of the two EDLs, respectively. It is important to note, however that the $\frac{1}{2}CV^2$ here is not purely electrostatic, with the entropic contribution contributing equally to the total free energy. For potentials beyond the steric threshold ($\Phi_{s\alpha} > \Phi_{ci}$), the total free energy exceeds $\frac{1}{2}CV^2$ due to the added contribution of the steric component. At the same time, when $\Phi_{s\alpha} \gg \Phi_{ci}$, the electrostatic component $F_{\mathrm{el}}$ and the steric component $F_{\mathrm{st}}$ are in the order of $\frac{1}{2}CV^2$ while the relative entropic contribution $F_{\mathrm{en}}$ is limited.

\begin{figure}[ht]
\centering
    \includegraphics[width=0.5\linewidth]{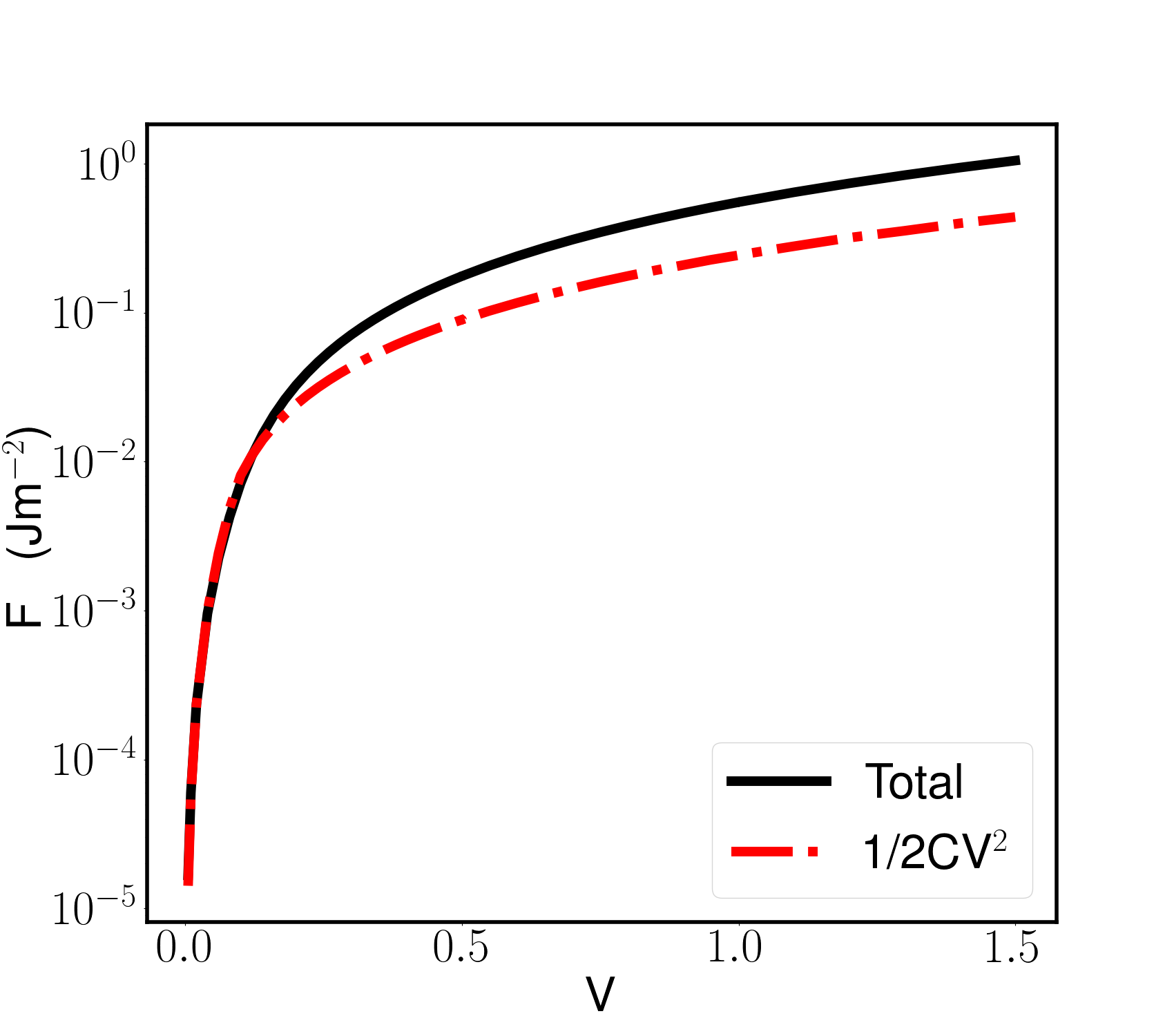}\hfil
    \includegraphics[width=0.5\linewidth]{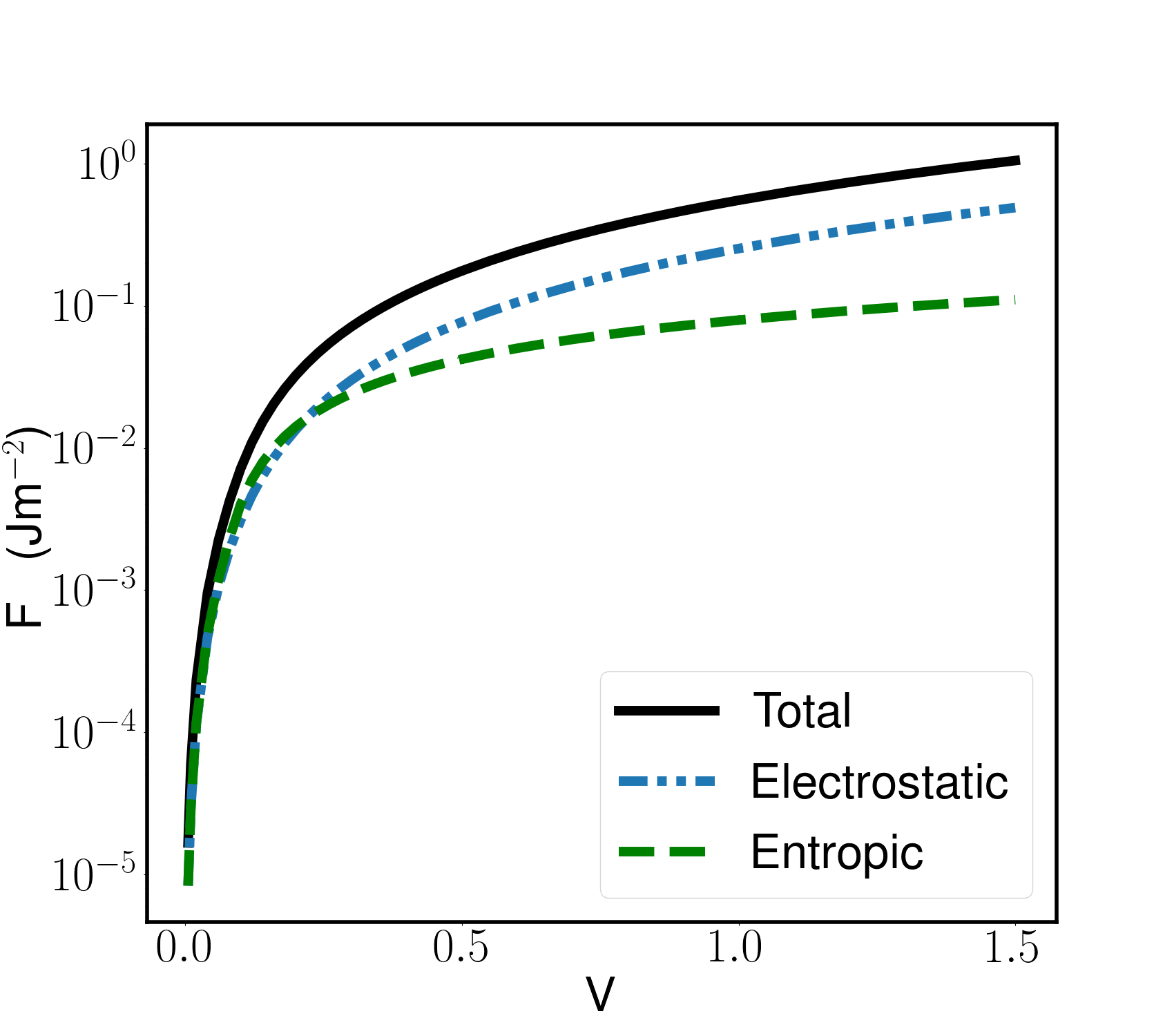}\par
    \includegraphics[width=0.5\linewidth]{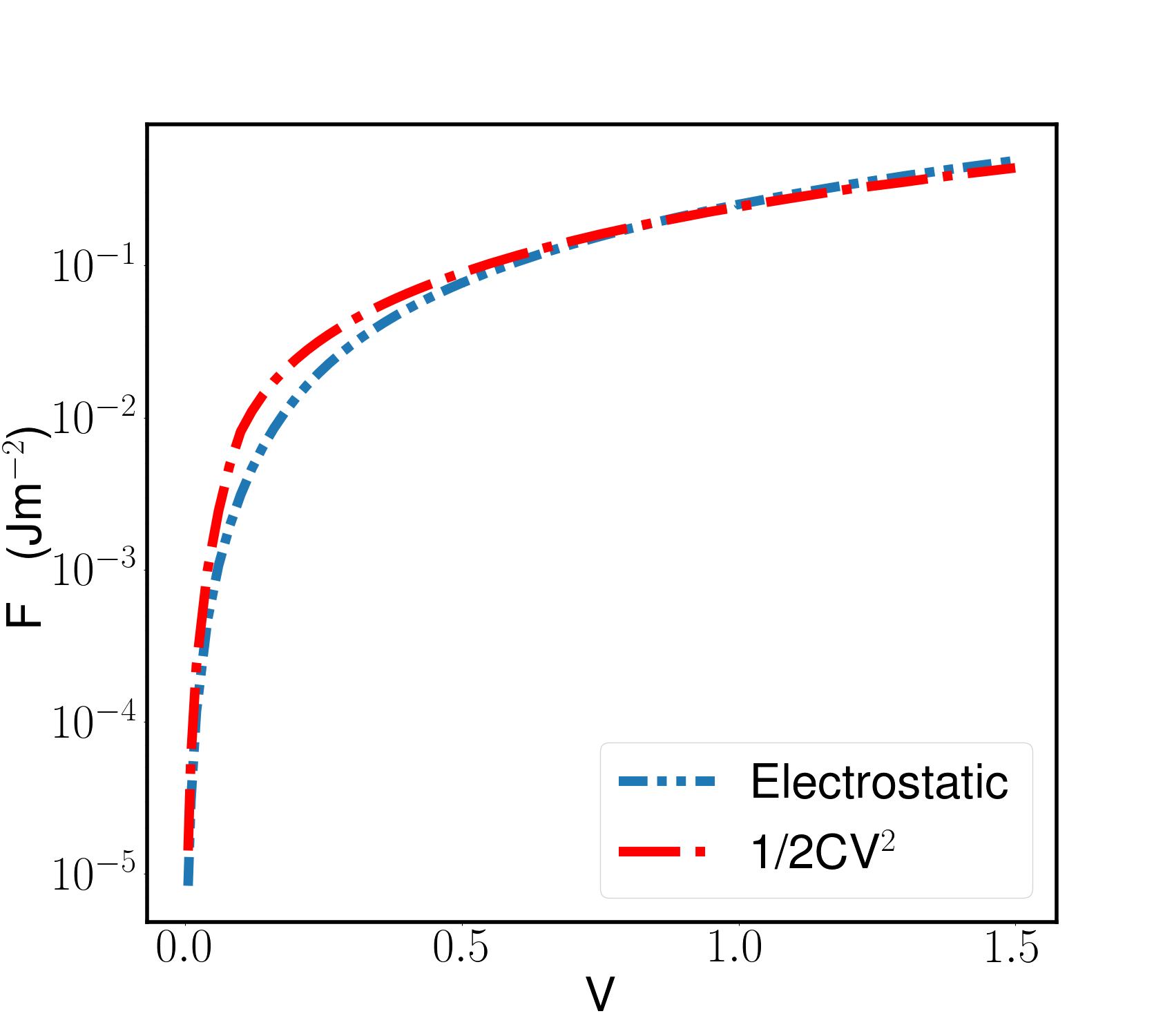}\hfil
    \includegraphics[width=0.5\linewidth]{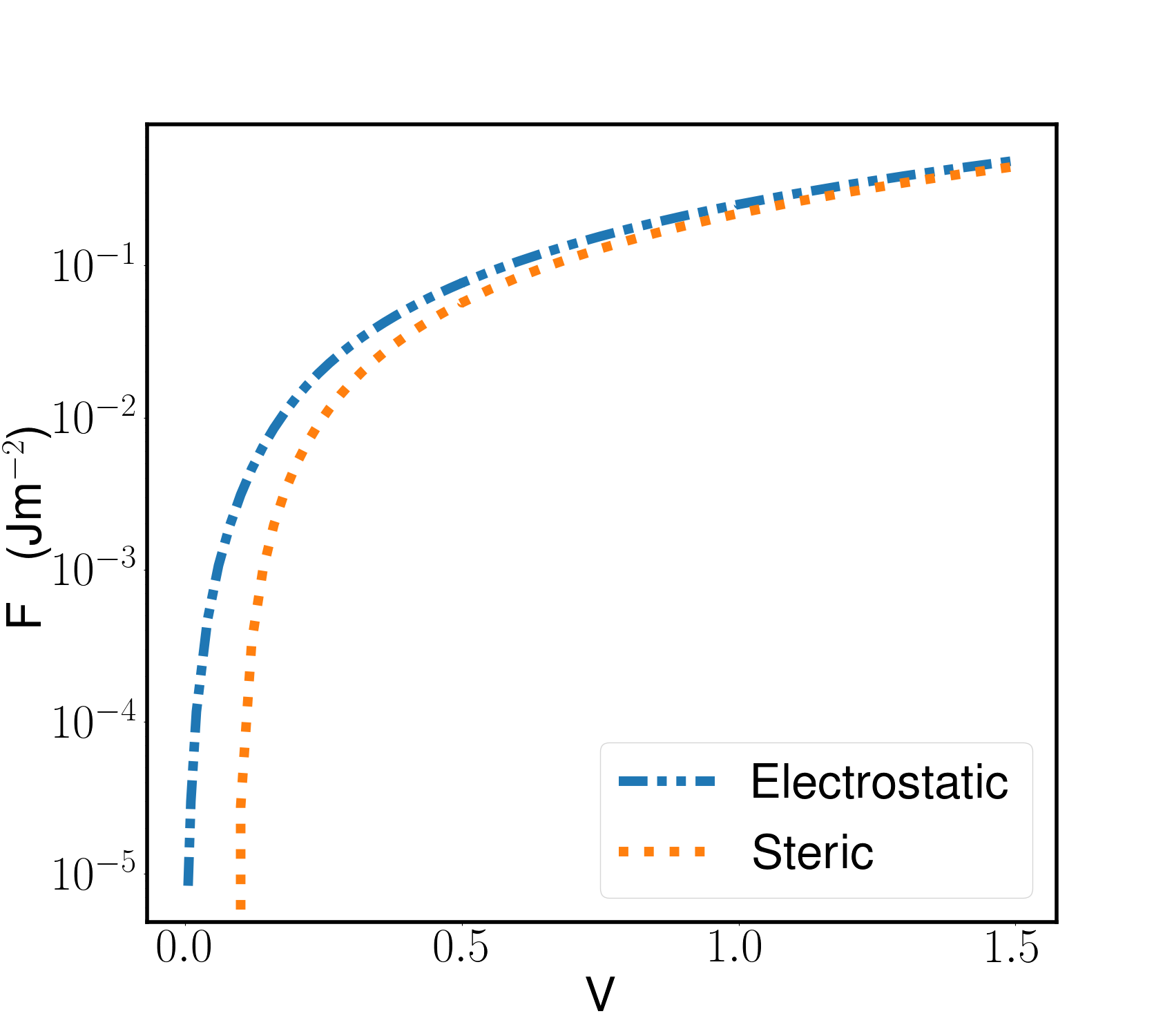}
\caption{Free energy components of LiTFSI. (Top left) total free energy curve against $\frac{1}{2}CV^2$. (Top right) total free energy curve against electrostatic component and $\frac{1}{2}CV^2$. (Bottom left) electrostatic energy component against $\frac{1}{2}CV^2$. (Bottom right) electrostatic energy component against steric and $\frac{1}{2}CV^2$.}\label{Fig:energyComp}
\end{figure}

The steric free energy is active above the steric threshold, with both the steric and electrostatic free energies increasing as potential difference increases. By contrast, up until the steric threshold, the entropic contribution increases with increasing potential difference, owing to the increasing concentration in the adsorption layer. Beyond the steric threshold, ion concentration caps lock the entropic contribution as shown in Fig.\ref{Fig:EneVsPD}(a), and its relative contribution to the total free energy declines as shown in Fig.\ref{Fig:EneVsPD}(b). 
\begin{figure}[ht]
\centering
    \includegraphics[width=0.5\linewidth]{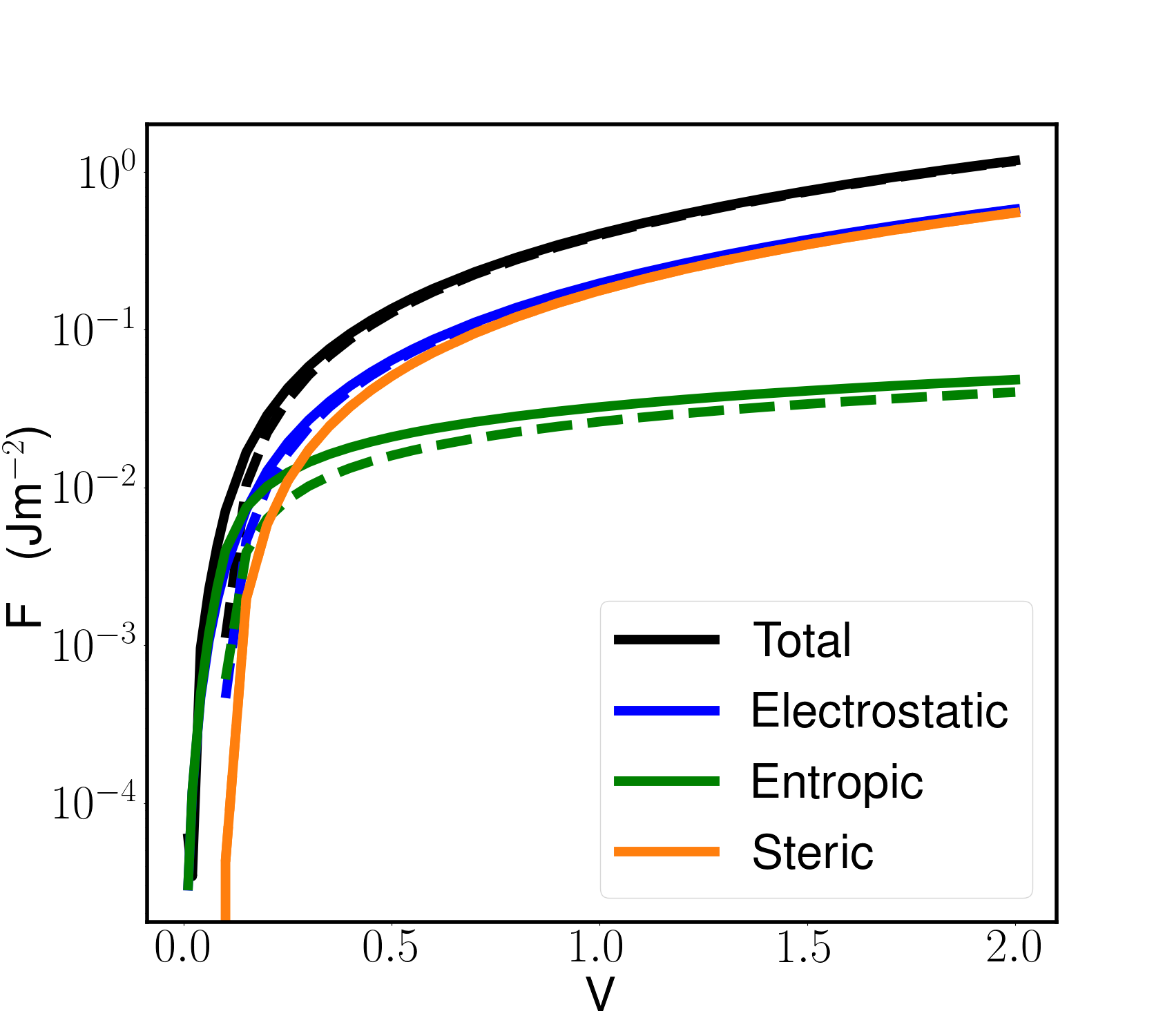}\hfil
    \includegraphics[width=0.5\linewidth]{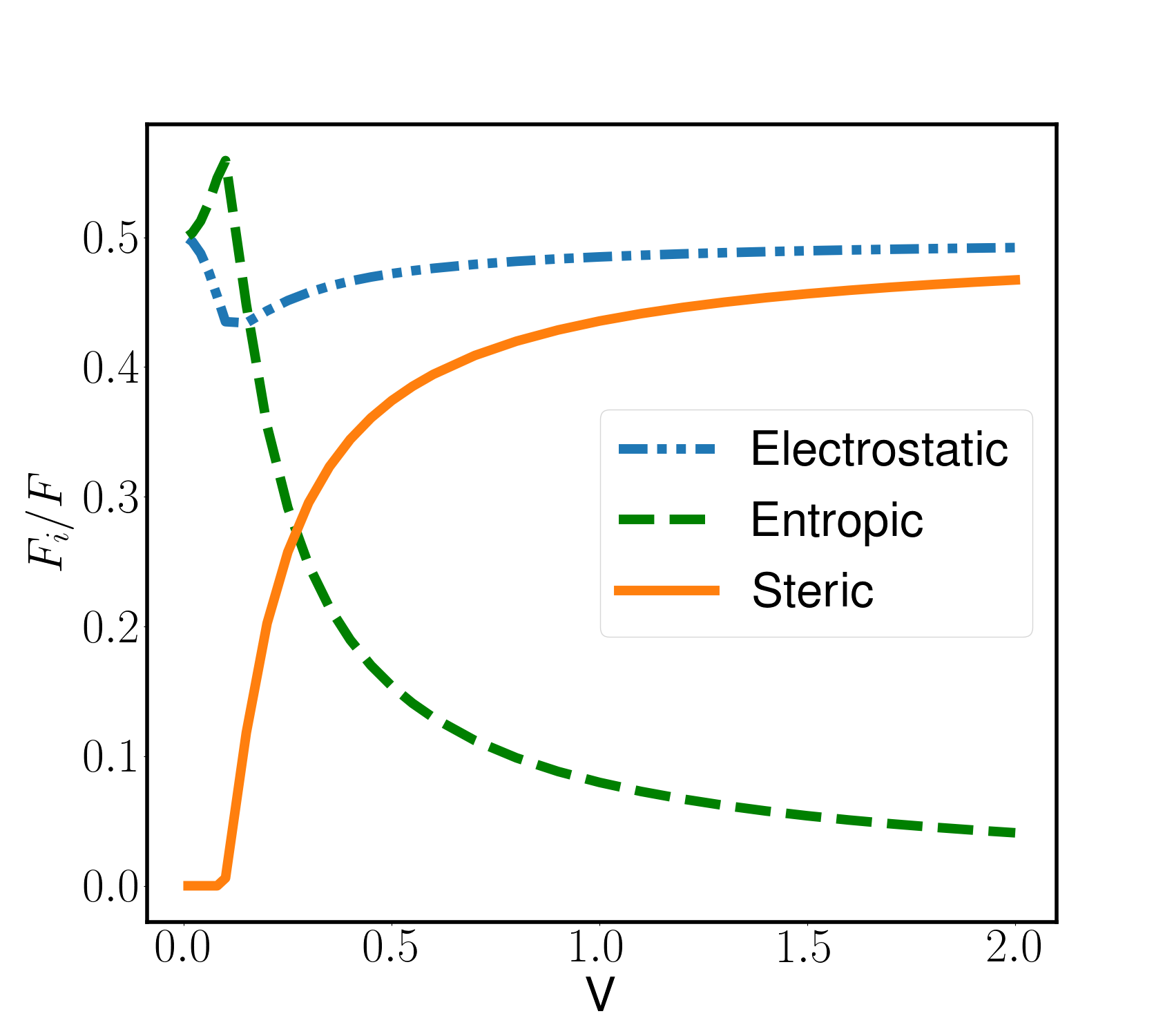}
\caption{(Left) Free energy components of \ch{EMTFSI}. Solid lines are values from the numerical solutions of equations \crefrange{eq:Fen}{eq:Fst}. The broken lines are the approximations given by \crefrange{eq:TFen}{eq:TFst}. (Right) The ratio of the components of the energy to the total energy. }
\label{Fig:EneVsPD}
\end{figure}
All the components of the free energy increase with increasing potential,  as shown in Fig. \ref{Fig:EneVsPD}(a).

The effect of the ion size on the total free energy has been investigated using the same electrolytes used in the calculation of capacitance. As expected, for potential below steric threshold the total free energy is independent of the ion size. For potentials above the steric limit, smaller ions provide a higher total free energy than larger ions. 

As stated above, the power relation between the Energy and the potential difference is approximately $E \propto V^2$ for potential below the threshold. However for potential above the threshold the power relation is ion-size specific, for small ions being close to 2 and decreases for larger ions to 1.44 being the power for largest ion size.
\begin{figure}[ht]
\centering
    \includegraphics[width=0.5\linewidth]{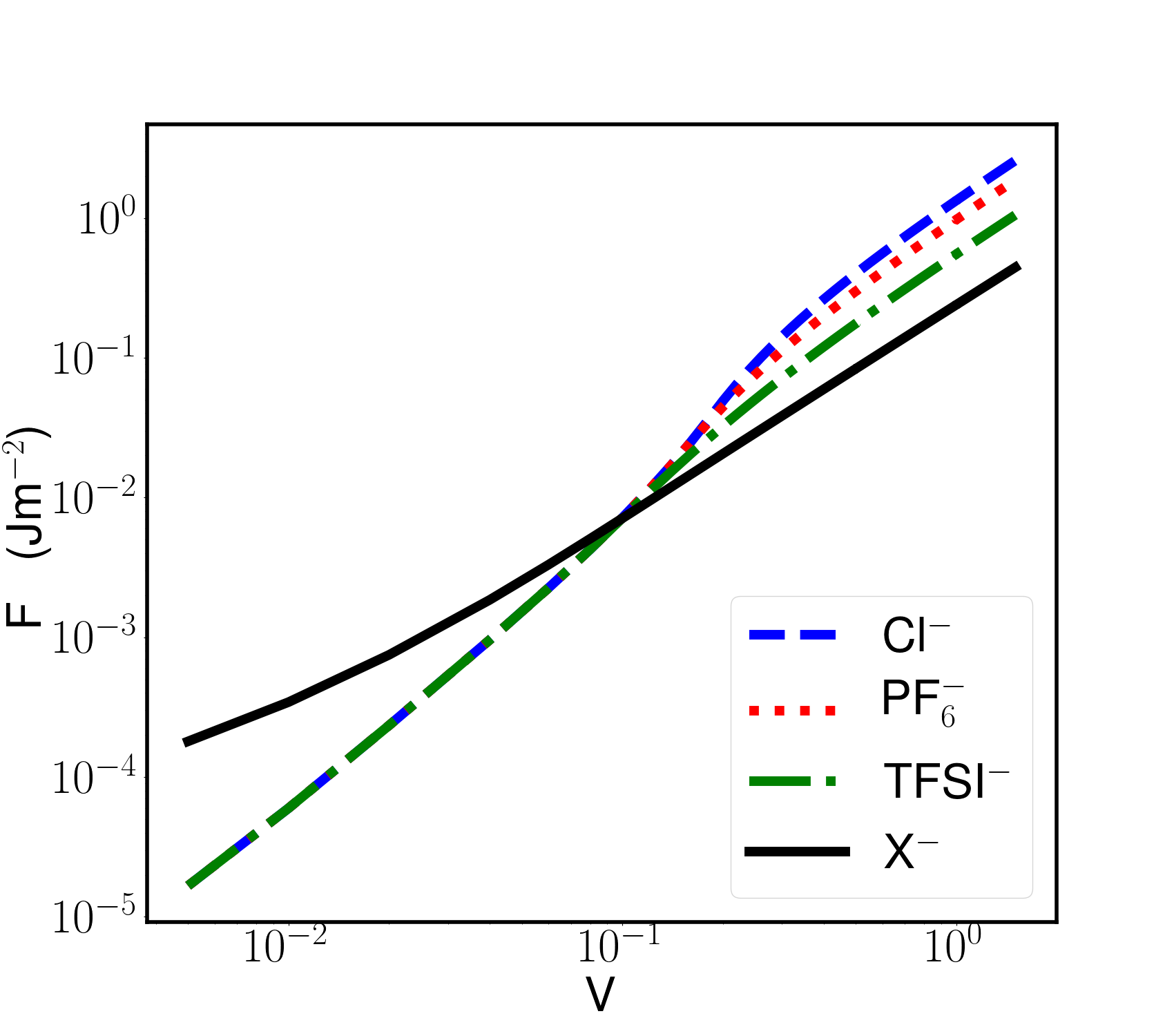}\hfil
    \includegraphics[width=0.5\linewidth]{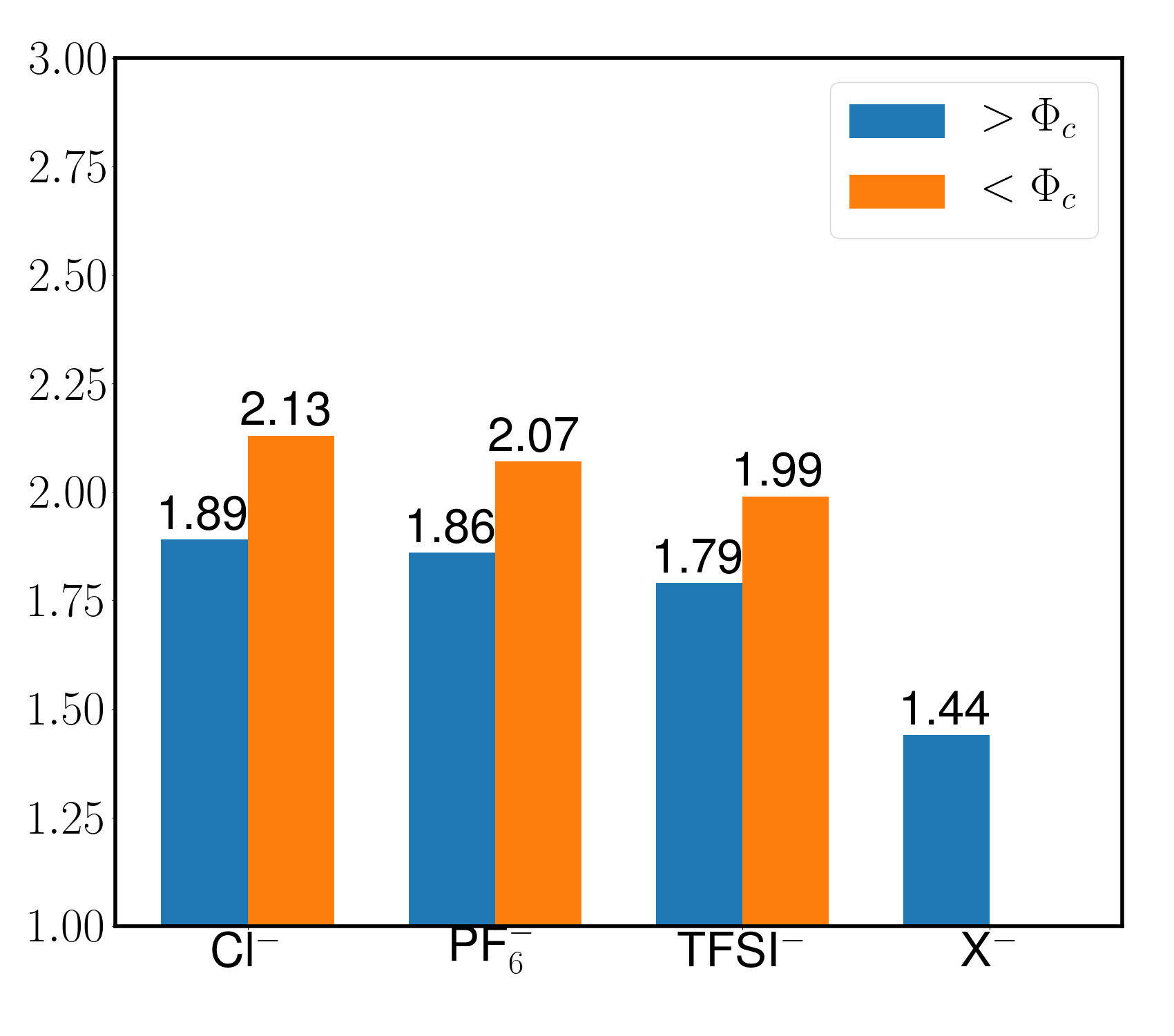}
\caption{(Left) Total free energy of the electrolytes against potential difference. (Right) Slope of the curves on the left before and after the steric threshold. }
\label{Fig:EneVsCap100NES}
\end{figure}

At the limit of high potential, the steric layers become very thick relative to the tail of the Debye length. Thus the EDL can be taken as a step function of concentration near the electrode, with concentration given by
\begin{equation}
    c_i(x)= 
\begin{cases}
    c_i^{\mathrm{cap}},& \mathrm{if } x\leq H_i\\
    c_{i\infty},              & \mathrm{otherwise}
\end{cases}
\end{equation}
For this charge distribution the corresponding electric field in the steric layer is
\begin{equation}
E_i(x) = \frac{\sigma_\alpha}{\varepsilon_0\varepsilon_s}+\frac{\rho_i^{\mathrm{cap}}}{\varepsilon_0\varepsilon_s}x
\end{equation}
\noindent where $\sigma_\alpha$ is the electrode surface charge density and $\rho_i^{\mathrm{cap}}=z_i \mathrm{e} N_{\mathrm{A}}c^{\mathrm{cap}}_i$ is the capped   volume charge density of the counterion.$z_i$ is the valence number of the counterion. The potential in the steric layer can be calculated from the electric field and is given by
\begin{equation}
    \Phi_i(x)=\Phi_{s\alpha}-\frac{\rho_i^{\mathrm{cap}}}{2\varepsilon_0\varepsilon_s}x^2-\frac{\sigma_\alpha}{\varepsilon_0\varepsilon_s}x
\end{equation}
$\Phi_{s\alpha}$ is the specific electrode potential.

\noindent Outside the steric layer, both the electric field and the potential are zero in the bulk. The corresponding free energy components for this high potential limiting case can be approximated for each electrode by
\begin{equation}\label{eq:TFen}
    F_{\mathrm{en}} = k_BTN_{\mathrm{A}}\sum_{i}\left[c_i^{\mathrm{cap}}\ln \frac{c_i^{\mathrm{cap}}}{c_{i\infty}}-c_i^{\mathrm{cap}} + c_{i\infty}\right]H_i
\end{equation}
\noindent for the entropic contribution. The electrostatic contribution is approximated by
\begin{equation}\label{eq:TFel}
F_{\mathrm{el}} = 
\frac{1}{2\varepsilon_0\varepsilon_s}\sum_i \biggl[ \frac{(\rho_i^{\mathrm{cap}})^2}{3} H_i^3 + \sigma_\alpha\rho_i^{\mathrm{cap}}H_i^2
+\sigma_\alpha^2H_i \biggr]
\end{equation}
Accordingly, the steric free energy is given by
\begin{equation}\label{eq:TFst}
\begin{split} F_{\mathrm{st}} &= 
\sum_i \biggl[ N_{\mathrm{A}}\mu_i^{\mathrm{cap}}c_i^{\mathrm{cap}} - \rho_i^{\mathrm{cap}}\left(\Phi_{s\alpha} -\frac{\sigma_\alpha H_i }{2\varepsilon_0\varepsilon_s} -\frac{\rho_i^{\mathrm{cap}}H_i^2}{6\varepsilon_0\varepsilon_s} \right) \biggr]H_i
\end{split}
\end{equation}
where $H_i$ is the steric layer thickness given by
\begin{eqnarray}
  H_i = \dfrac{z_i\sigma_\alpha}{\rho_i^{\mathrm{cap}}}\Bigg[1 - \sqrt{1-\dfrac{2\varepsilon_0\varepsilon_s\rho_i^{\mathrm{cap}}}{\sigma_\alpha^2}(\Phi_{ci} - \Phi_{s\alpha})}\Bigg]
\end{eqnarray}
For $z$:$z$ symmetric electrolytes $H_i$, and the electrode charge density, $\sigma_\alpha$, are given by \cite{PhysRevE.75.021502}
\begin{eqnarray}\label{SLT}
  \nonumber  H_i &=&\lambda_D\sqrt{2\nu_i}\Bigg[-1+\frac{\nu_i}{2}+\sqrt{\left(1-\frac{\nu_i}{2}\right)^2-\frac{z_i\mathrm{e}\Phi_{s\alpha}}{k_BT}-\ln{\frac{2}{\nu_i}}}\Bigg]\\
  \sigma_\alpha&=& -2\rho_i^{\mathrm{bulk}}\lambda_D\sqrt{\frac{2}{\nu_i}\Bigg[\left(1-\frac{\nu_i}{2}\right)^2-\frac{z_i\mathrm{e}\Phi_{s\alpha}}{k_BT}-\ln{\frac{2}{\nu_i}}\Bigg]}
\end{eqnarray}
\noindent where $\nu_i = 2c_{i\infty}/c^{\mathrm{cap}}_i$ and $\rho_i^{\mathrm{bulk}} =z_i \mathrm{e} N_{\mathrm{A}}c_{i\infty}$ is volume charge density at bulk.

\crefrange{eq:TFen}{eq:TFst} are approximations of \crefrange{eq:Fen}{eq:Fst}, respectively valid for potentials beyond the steric limit. The approximations converge to the theoretical values at higher voltages as shown in Fig.\ref{Fig:EneVsPD}(a) with relative error less than 1\% for the total free energy at 2V.

In conclusion, we introduced steric terms in the excess chemical potential of an ion to account for size of an ion that induces an ion concentration cap. Concentration profiles, charge densities and differential capacitance are similar to that of the composite diffuse-layer model. The added steric chemical potential was used as a short range ion-ion interaction represent ion-ion correlation effects and is included in the total free energy. We have shown that $\frac{1}{2}CV^2$ is an oversimplified calculation of the energy of an EDL. We have also derived an analytical approximation to predict the free energies in terms of the given parameters, describing behaviour where the steric interaction becomes active (when $V > 0.2$V). 



\bibliography{library}

\end{document}